\documentclass[conference]{IEEEtran}
\IEEEoverridecommandlockouts
\usepackage{cite}
\usepackage{amsmath,amssymb,amsfonts}
\usepackage{algorithm, algpseudocode}
\usepackage{graphicx}
\usepackage{textcomp}
\usepackage{xcolor}
\usepackage[utf8]{inputenc}
\usepackage{kotex}
\usepackage{kotex-logo}

\def\BibTeX{{\rm B\kern-.05em{\sc i\kern-.025em b}\kern-.08em
    T\kern-.1667em\lower.7ex\hbox{E}\kern-.125emX}}

\begin{document}

\title{3D-EDM: Early Detection Model for \\ 3D-Printer Faults}

\author{\IEEEauthorblockN{Harim Jeong}
\IEEEauthorblockA{\textit{Department of Cultural Management} \\
\textit{Sungkyunkwan University}\\
Seoul, South Korea \\
harim4110@gmail.com}
\and
\IEEEauthorblockN{Joo Hun Yoo}
\IEEEauthorblockA{\textit{Department of Artificial Intelligence} \\
\textit{Sungkyunkwan University}\\
Suwon, South Korea \\
andrewyoo@g.skku.edu}
}

\maketitle

\begin{abstract}
With the advent of 3D printers in different price ranges and sizes, they are no longer just for professionals. However, it is still challenging to use a 3D printer perfectly. Especially, in the case of the Fused Deposition Method, it is very difficult to perform with accurate calibration. Previous studies have suggested that these problems can be detected using sensor data and image data with machine learning methods. However, there are difficulties to apply the proposed method due to extra installation of additional sensors. Considering actual use in the future, we focus on generating the lightweight early detection model with easily collectable data. Proposed early detection model through Convolutional Neural Network shows significant fault classification accuracy with 96.72\% for the binary classification task, and 93.38\% for multi-classification task respectively. By this research, we hope that general users of 3D printers can use the printer  accurately. 
\end{abstract}


%%%%%%%%%%%%%%%%%%%%%%%%%%%%%%%%%%%%%%%%%%%%%%%%%%
%%%%%%%%%%%%%%%%%%%%%%%%%%%%%%%%%%%%%%%%%%%%%%%%%%
\section{Introduction} \label{sec:intr}

3D printing technology, which began in the early 2000s, continues to develop at a rapid pace. Since its emergence, the technology has been extended to various industries. Initially, it was limited for professional usages, but recently, inexpensive printers for hobbies have been appeared, and the market has grown to general-purpose. For those who use 3D printer as a hobby, Fused Deposition Modeling (FDM) type printers are mainly used rather than Stereolithography (SLA) or Digital light processing (DLP)type printers.

To perform the FDM type of 3D printer, the calibration process is required depending on the surrounding temperature, bed types, size of the nozzles, and type of filaments. Compared to a resin printer that uses only a z-axis motor and display, the FDM method contains at least four stepper motors, one for each x, y, z-axis, and extruder. In addition, rails, belts, nozzles, and fans are making the structure more complex. It is challenging for the users to perform and maintain perfect calibration in such environments.

Faults that often occur on the 3D printer are typically layer shifts, strings, warping, and under extrusion as shown in Fig.1. Printing each 3-dimensional object takes a long time, and it is hardly possible to watch and modify the settings. Therefore, even if a significant defect occurs, it is difficult to immediately stop and re-calibrate. This is the reason why real-time fault detection or early detection is necessary for 3D printing.

In this study, the 3D printer Early Detection Model (3D-EDM) is proposed as lightweight and high performance, using the image deep learning techniques.

%%%%%%%%%%%%%%%%%%%%%%%%%%%%%%%%%%%%%%%%%%%%%%%%%%
%%%%%%%%%%%%%%%%%%%%%%%%%%%%%%%%%%%%%%%%%%%%%%%%%%

\begin{figure}[H]
\centering
\includegraphics[width=0.90\linewidth]{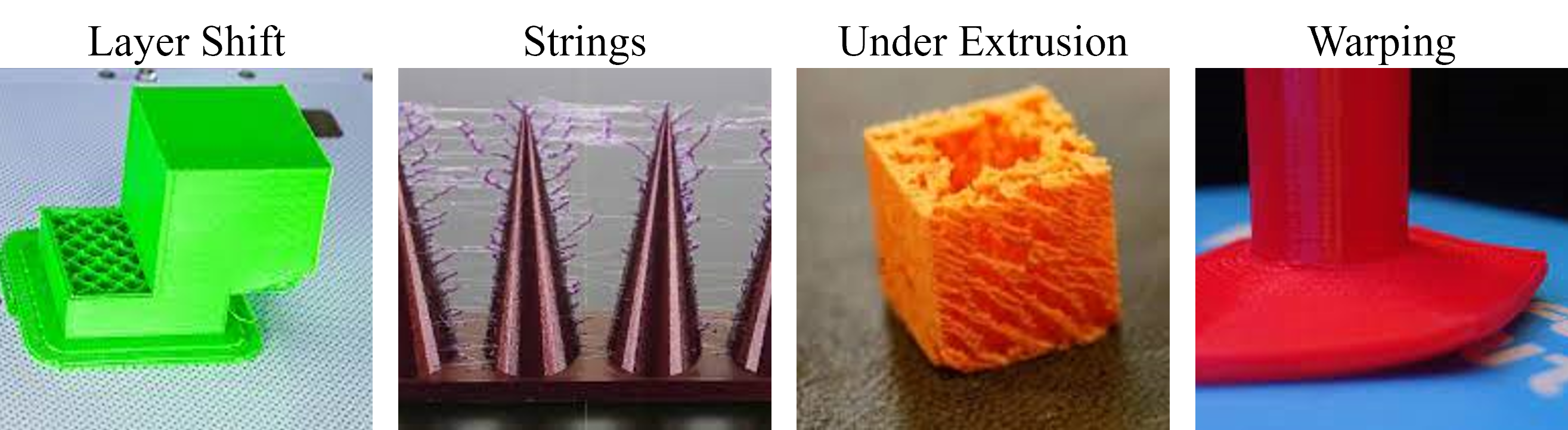}
\centering
\caption{Sample images of 3D printer faults}
\end{figure}

\section{Fault Detection in 3D Printer} \label{sec:2}

In order to reduce the time and economic waste of 3D printers, fault detection has been applied in various ways. Many researchers have attempted to obtain information on defects by analyzing the various dataset collected from printers.

Banadaki suggests a fault detection method through additive manufacturing with the speed and temperature of the extruder \cite{banadaki2020toward}. However, since the data available from the printer itself is limited, several studies conducted experiments using additional sensors such as vibrarion sensor. Bing  proposed a real-time detection solution with Support Vector Machine (SVM) classifier in \cite{li20193d}. With additional vibration sensor, they effectively detect the failures of 3D printers.

In addition to the method using the sensor data, there are also studies that detects failures of the 3D printer with the printing photos collected through a camera. In the case of Delli's research \cite{delli2018automated}, the RGB value of each critical checkpoint of the object model are evaluated. They monitor whether the output was normally performed by comparing the RGB values of the photo taken at the checkpoint during actual printing process. Kadam, on the other hand, suggests a detection method through machine learning algorithms on the first layer’s top image \cite{kadam2021enhancing}.  Referring the importance of the first layer, they compared the classification results from following pre-trained models; EfficientnetB0, Resnet18, Resnet50, Alexnet, and Googlenet. Finally, Jin proved that by attaching a camera right next to the nozzle to recognize the nozzle edge, the CNN-based classifiers could be used to classify whether printing is performed correctly in real-time \cite{jin2020automated}.

These previous studies have proven that various 3D printing failures can be diagnosed, detected, and predicted. Still, there may be difficulties in actual use. A camera must be mounted on the nozzle head to see the top view of the printing, but increasing the total weight by mounting a camera on a nozzle can cause another failure. Based on the related studies, our paper aims to conduct research focusing on practical use.

%%%%%%%%%%%%%%%%%%%%%%%%%%%%%%%%%%%%%%%%%%%%%%%%%%
%%%%%%%%%%%%%%%%%%%%%%%%%%%%%%%%%%%%%%%%%%%%%%%%%%
\section{Early Detection Model} \label{sec:3}

\subsection{Dataset}

In this research, training images were crawled from Google and YouTube to collect massive real-world data. The data collection using crawling was to cover various printing environments, including backgrounds, filament colors, and the shape of objects. 

To detect the incorrect output and confirm the type of faults, 200 images were collected for each layer shift, strings, under extrusion, and warping cases. About 500 normally printed data were also gathered. Accordingly, a total of about 1.3k image data was selected for the experiment.

\subsection{Experiment Settings}

To perform an image deep learning for the fault detection, we preprocess collected datasets with various data transformation tools. Since the collected data all have different sizes, it was set to focus on the object using the resize and centercrop functions. The vertical flip and horizontal flip functions were also applied to train the model to effectively recognize  the images taken from a different angle.

Training model that can detect defects based on the classifier, the Convolutional Neural Network (CNN) model  was adopted. There are already fine pre-trained image classification models, our experiments were focused to generate lighter models. After generating a training model with 10 convolutional layers, we tried to find an optimization model in the form of removing one layer at a time. Experiment was conducted only when the test performance of the model exceeded 90.0\%. Final model is shown in Fig. 2.

\begin{figure}[!ht]
\centering
\includegraphics[width=0.50\linewidth]{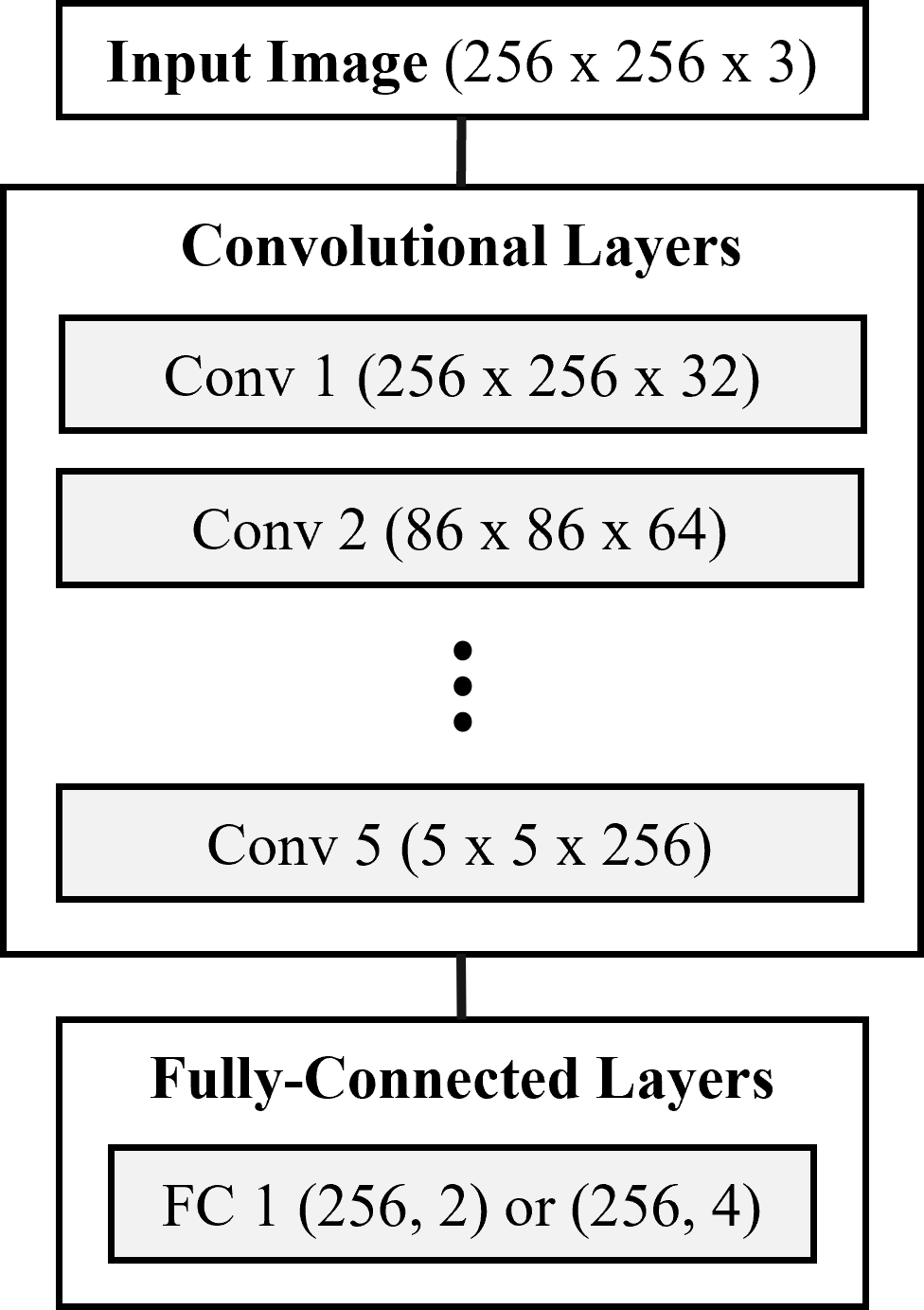}
\centering
\caption{Sample images of 3D printer faults}
\end{figure}

The size of 256 x 256 preprocessed image data is used as an input. It passes through a total of five convolutional layers with attached pooling layers to train the significant features from each part of the images. After that, the training ends with two or four output nodes through one fully-connected layer. As mentioned above, setting two output nodes determines whether the input value is the fault result. Setting four output nodes tells the user in detail what issues the incorrect result contains respectively.

\subsection{Experiment Results}

Since no other studies were experimented from various angles for all possible failure cases as in this study, a performance comparison was made with studies with high test accuracy or similar experimental processes. Table 1 contains the accuracy comparison results.

\begin{table}[h]
\caption{Accuracy comparison with other studies}
\centering
\begin{tabular}{|c|c|c|}
    \hline
    % \rowcolor[HTML]{C0C0C0} 
    Research                                                & Analysis                                                                                                             & Acc     \\ \hline
    {[}2{]}                                                 & \begin{tabular}[c]{@{}c@{}}SVM on vibration sensor \\ data to detect failures\end{tabular}                           & 94.93\% \\ \hline
    {[}5{]}                                                 & \begin{tabular}[c]{@{}c@{}}CNN classifier on surface images \\ to detect warping failures\end{tabular}               & 91.0\%  \\ \hline
    \begin{tabular}[c]{@{}c@{}}3D-EDM\\ Binary\end{tabular} & \begin{tabular}[c]{@{}c@{}}CNN classifiers on\\ multi-angle printed images\\ decide fault or not\end{tabular}        & 96.72\% \\ \hline
    \begin{tabular}[c]{@{}c@{}}3D-EDM\\ Multi\end{tabular}  & \begin{tabular}[c]{@{}c@{}}CNN classifiers on 4 specific\\ fault cases to distinguish\\ the fault cases\end{tabular} & 93.38\% \\ \hline
\end{tabular}
\end{table}

Study with SVM classifiers on various sensors \cite{li20193d}, especially with vibration sensors, shows 94.93\% of test accuracy for detecting failure printed objects. Detection algorithm with CNN classifier on surface images showed 91.0\% test accuracy \cite{jin2020automated}. Although these studies show a high detection accuracy of more than 90\%, as mentioned above, there are remaining issues for actual 3D-printer users to utilize the solutions.

We seperated the test results into two significant directions and showed the results for each experiment.  In the case of binary-classification, which uses 3D-EDM to check whether the received input data has a fault or not, the test accuracy was 96.72\%. In addition, the multi-classification task, that distinguishes the current fault case from layer shift, strings, under extrusion, and warping, showed about 93.38\% of the test accuracy.

%%%%%%%%%%%%%%%%%%%%%%%%%%%%%%%%%%%%%%%%%%%%%%%%%%
%%%%%%%%%%%%%%%%%%%%%%%%%%%%%%%%%%%%%%%%%%%%%%%%%%

\section{Conclusion}

In this study, we generated a lightweight image deep learning algorithm, to predict and detect possible fault cases in a 3D printing environment. Although previous studies have suggested a number of methods for fault detection, our study is distinct in that we designed the experiment for practical use. In addition, we believe it is more suitable for actual use because it utilizes easily obtainable usage data rather than data that from complicated installation settings.

As with many deep learning-related models, this study could show more accurate results if more data were collected, but there was a time limitation to collect more data at present.  When more data collected, it will be possible to make a more accurate Early Detection Model. However, this study is still meaningful because it has been proven that this method is feasible, and optimization for this method has been proposed. In the near future, we hope that by running this model on Raspberry Pi for actual use, 3D printer users can print the desired object as they want.

\bibliographystyle{IEEEtran}
\bibliography{biblio}

%\begin{thebibliography}{00}
%\end{thebibliography}
%\vspace{12pt}

\end{document}